\pdfoutput=1

\documentclass[11pt,twoside,a4paper,cmspaper,final,collab]{cms-tdr}
\def\svnVersion{294870}\def\cmsCernNo{2013-037}\def\cmsCernDate{\today}\def\cmsMessage{Published in Physical Review Letters as \href{http://dx.doi.org/10.1103/PhysRevLett.115.012301}{\doi{10.1103/PhysRevLett.115.012301.}}}

\begin{document}\cmsNoteHeader{HIN-14-006}

\hyphenation{had-ron-i-za-tion}
\hyphenation{cal-or-i-me-ter}
\hyphenation{de-vices}
\RCS$Revision: 284910 $
\RCS$HeadURL: svn+ssh://svn.cern.ch/reps/tdr2/papers/HIN-14-006/trunk/HIN-14-006.tex $
\RCS$Id: HIN-14-006.tex 284910 2015-04-16 20:06:39Z qwang $
\newlength\cmsFigWidth
\ifthenelse{\boolean{cms@external}}{\setlength\cmsFigWidth{0.98\columnwidth}}{\setlength\cmsFigWidth{0.75\textwidth}}
\ifthenelse{\boolean{cms@external}}{\providecommand{\cmsLeft}{top}}{\providecommand{\cmsLeft}{left}}
\ifthenelse{\boolean{cms@external}}{\providecommand{\cmsRight}{bottom}}{\providecommand{\cmsRight}{right}}
\providecommand{\rootsNN}{\ensuremath{\sqrt{s_{_\mathrm{NN}}}}\xspace}
\providecommand{\roots}{\ensuremath{\sqrt{s}}\xspace}
\providecommand{\dAu}{\ensuremath{\text{dAu}}\xspace}
\providecommand{\PbPb}{\ensuremath{\text{PbPb}}\xspace}
\providecommand{\pPb}{\ensuremath{\text{pPb}}\xspace}
\providecommand{\pp}{\ensuremath{\text{pp}}\xspace}
\providecommand{\EPOS}{\textsc{epos}\xspace}
\newcommand{\dmean}[1]{\ensuremath{\langle\langle#1\rangle\rangle}}
\newcommand{\cn}[1]{\ensuremath{c_n\{#1\}}}
\newcommand{\ctwo}[1]{\ensuremath{c_2\{#1\}}}
\newcommand{\vn}[1]{\ensuremath{v_n\{#1\}}}
\newcommand{\vtwo}[1]{\ensuremath{v_2\{#1\}}}
\newcommand{\noff}{\ensuremath{N_\text{trk}^\text{offline}}\xspace}
\newcommand{\dphi}{\ensuremath{\Delta\phi}}

\cmsNoteHeader{HIN-14-006}
\title{Evidence for collective multi-particle correlations in \pPb collisions}

\date{\today}

\abstract{
The second-order azimuthal anisotropy Fourier harmonics, $v_2$, are obtained in \pPb and \PbPb collisions
over a wide pseudorapidity ($\eta$) range based on correlations among six or more charged particles.
The \pPb data, corresponding to an integrated
luminosity of 35\nbinv, were collected during the 2013
LHC \pPb run at a nucleon-nucleon center-of-mass energy of 5.02\TeV by the CMS experiment.
A sample of semi-peripheral \PbPb collision data at $\rootsNN= 2.76$\TeV, corresponding to an integrated luminosity of 2.5\mubinv
and covering a similar range of particle multiplicities as the \pPb data,
is also analyzed for comparison.
The six- and eight-particle cumulant and the Lee-Yang zeros  methods are used to
extract the $v_2$ coefficients,
extending previous studies of two- and four-particle correlations.
For both the \pPb and \PbPb systems,
the $v_2$ values obtained with correlations among more than four particles
are consistent with previously published four-particle results.
These data support the interpretation of a collective origin for the previously observed long-range (large $\Delta\eta$) correlations in both systems.
The ratios of $v_2$ values corresponding to correlations including different numbers of particles are compared to
theoretical predictions that assume a hydrodynamic behavior
of a \pPb system dominated by fluctuations in the positions of participant nucleons.
These results provide new insights into the multi-particle dynamics of collision systems with a very small overlapping region.
}

\hypersetup{%
pdfauthor={CMS Collaboration},%
pdftitle={Evidence for collective multi-particle correlations in pPb collisions},%
pdfsubject={CMS},%
pdfkeywords={CMS, physics, heavy-ion, correlation}}

\maketitle

Measurements at the CERN LHC have led to the discovery of
two-particle azimuthal correlation structures at large relative pseudorapidity (long-range)
in proton-proton (\pp)~\cite{Khachatryan:2010gv} and
proton-lead (\pPb)~\cite{CMS:2012qk,alice:2012qe,atlas:2012fa,Abelev:2014mda} collisions.
Similar long-range structure has also been observed for $\rootsNN = 200$\GeV
deuteron-gold (\dAu) collisions
at RHIC~\cite{Adare:2013piz,Adare:2014keg}.
The results extend previous studies of relativistic heavy ion collisions,
such as for the copper-copper ~\cite{Alver:2008gk}, gold-gold ~\cite{Adams:2005ph,Abelev:2009af,Alver:2008gk,Alver:2009id,Abelev:2009jv},
and lead-lead (\PbPb)~\cite{Chatrchyan:2011eka,Chatrchyan:2012wg,Aamodt:2011by,ATLAS:2012at,Chatrchyan:2012ta,CMS:2013bza} systems,
where similar long-range, two-particle correlations
at small relative azimuthal angle $\abs{\dphi} \approx 0$ were first observed.
A fundamental question is whether the observed behavior results from correlations exclusively
between particle pairs, or if it is
a multi-particle, collective effect. It has been suggested that the hydrodynamic collective flow of a strongly interacting and expanding medium~\cite{Ollitrault:1992bk,Heinz:2013th,Gale:2013da} is responsible for these long-range correlations in central and mid-central heavy ion collisions.
The origin of the observed long-range correlations in collision systems with a small overlapping region, such as for \pp and \pPb collisions, is not clear since
for these systems the formation of an extended hot medium is not necessarily expected.
Various theoretical models have been proposed to interpret the \pp \cite{Li:2012hc,Bjorken:2013boa} and \pPb results,
including initial-state gluon saturation without any final state interactions~\cite{Dusling:2012wy,Dusling:2012cg} and, similar to what is thought to occur in heavier systems, hydrodynamic behavior that develops in a conjectured
high-density medium ~\cite{Schenke:2014zha,Bozek:2011if,Bozek:2012gr}.
These models have been successful in describing different aspects of the previous experimental results.

To further investigate the multi-particle nature of the observed
long-range correlation phenomena, in this Letter we present
measurements of correlations among six or more charged particles for \pPb collisions at a center-of-mass energy per nucleon pair of $\rootsNN = 5.02$\TeV.
The azimuthal dependence of particle production is
typically characterized by an expansion in Fourier harmonics ($v_n$)~\cite{Voloshin:1994mz}.
In hydrodynamic models, the second ($v_2$) and third ($v_3$) harmonics, called ``elliptic" and ``triangular'' flow~\cite{Alver:2010gr},
respectively, directly reflect the response to the initial collision geometry and fluctuations~\cite{Alver:2010dn,Schenke:2010rr,Qiu:2011hf}, providing insight into the fundamental transport
properties of the medium.
First attempts to establish the multi-particle nature of the correlations observed in \pPb collisions were presented in Refs.~\cite{Aad:2013fja,Chatrchyan:2013nka}
by directly measuring four-particle azimuthal correlations, where the elliptic flow signal was obtained using the four-particle cumulant method~\cite{Bilandzic:2010jr}.
However, four-particle correlations can still be affected by contributions from non-collective effects such as fragmentation of back-to-back jets. By extending the studies to six- and eight-particle cumulants~\cite{Bilandzic:2010jr} and by  also obtaining results using the Lee-Yang zeros (LYZ) method, which involves correlations among all detected particles~\cite{NuclPhysA.727.373, Borghini:2004ke}, it is possible to further explore the collective nature of the correlations. High-statistics data obtained
by the CMS experiment during the 2013 \pPb run at the LHC are used.
With a sample of very high final state multiplicity \pPb collisions,
the correlation data have been studied in a regime that is comparable to the charged particle multiplicity of the 50\% most peripheral (semi-peripheral) \PbPb collisions at $\rootsNN = 2.76$\TeV.

The CMS detector comprises a number of subsystems~\cite{JINST}.
The results in this Letter are mainly based on the
silicon tracker information. The silicon tracker, located in the 3.8\unit{T} field of a
superconducting solenoid, consists of 1\,440 silicon pixel and 15\,148 silicon strip
detector modules. The silicon tracker measures charged particles within the pseudorapidity
range $\abs{\eta}< 2.5$, and provides an impact parameter resolution of ${\approx}15\mum$ and
a transverse momentum (\pt) resolution better than 1.5\% at $\pt \approx 100\GeVc$.
The electromagnetic (ECAL) and hadron (HCAL) calorimeters are also
located inside the solenoid and cover the pseudorapidity range $\abs{\eta} < 3.0$.
The HCAL barrel and endcaps are sampling calorimeters composed of brass and
scintillator plates.
The ECAL consists of lead tungstate crystals arranged in a quasiprojective geometry.
Iron/quartz-fiber \v Cerenkov hadron forward (HF) calorimeters cover the range $2.9 < \abs{\eta} < 5.2$ on either
side of the interaction region.
These HF calorimeters are azimuthally subdivided into $20^{\circ}$
modular wedges and further segmented to form $0.175\times0.175$ rad $(\Delta\eta \times \Delta\phi)$ ``towers".
The detailed Monte Carlo (MC) simulation of the CMS detector response is based
on {\GEANTfour} \cite{GEANT4}.

The analysis is performed using data recorded by CMS during the
LHC \pPb run in 2013. The data set corresponds
to an integrated luminosity of 35\nbinv.
The beam energies were 4\TeV for protons and 1.58\TeV
per nucleon for lead nuclei, resulting in
\rootsNN\ = 5.02\TeV. The beam directions were reversed during the run allowing a check of one potential source of systematic uncertainties.
As a result of the energy difference between the colliding beams, the nucleon-nucleon
center-of-mass in the \pPb collisions is not at rest with respect to the laboratory frame.
Massless particles emitted at $\eta_\text{cm} = 0$ in the nucleon-nucleon
center-of-mass frame will be detected at $\eta = -0.465$ (clockwise proton beam)
or 0.465 (counterclockwise proton beam) in the laboratory frame.
A sample of $\rootsNN = 2.76$\TeV \PbPb data collected during the 2011 LHC heavy-ion run,
corresponding to an integrated luminosity of 2.3\mubinv, is also analyzed for
comparison purposes. The triggers and event selection, as well as track reconstruction
and selection are summarized below and are identical to those used in Ref.~\cite{Chatrchyan:2013nka}.

Minimum bias (MB) \pPb events were triggered by requiring at least
one track with $\pt > 0.4$\GeVc to be found in the pixel tracker
for a \pPb bunch crossing.
Only a small fraction (${\sim}10^{-3}$) of all MB
triggered events were recorded (\ie, the trigger was ``prescaled'')
because of hardware limits on the data acquisition rate.
In order to select high-multiplicity \pPb collisions,
a dedicated high-multiplicity trigger was implemented using the CMS level-1
(L1) and high-level trigger (HLT) systems.
At L1, three triggers requiring the total transverse energy summed over ECAL and HCAL
to be greater than 20, 40, and 60\GeV were used since these cuts selected roughly the same events as the three HLT multiplicity selections discussed below.
Online track reconstruction for the HLT was based on
the three layers of pixel detectors, and required a track origin within a cylindrical region of
length 30\unit{cm} along the beam and radius 0.2\unit{cm} perpendicular to the beam around the nominal interaction point.
For each event, the vertex reconstructed with the highest number of pixel tracks was selected.
The number of pixel tracks (${N}_\text{trk}^\text{online}$)
with $\abs{\eta}<2.4$, $\pt > 0.4\GeVc$, and a distance of closest approach to this vertex of 0.4\unit{cm} or
less, was determined for each event. Several high-multiplicity ranges were defined with
prescale factors that were progressively reduced until, for the highest multiplicity events, no prescaling was applied.

In the offline analysis, hadronic collisions are selected by requiring a
coincidence of at least one HF calorimeter tower containing more than 3\GeV of
total energy in each of the HF detectors.
Only towers within $3<\abs{\eta}<5$ are used to avoid the edges of the HF acceptance.
Events are also required to contain at least one reconstructed primary
vertex within 15\unit{cm} of the nominal interaction point along the beam axis
and within 0.15\unit{cm} transverse to the beam trajectory.
At least two reconstructed tracks are required to be associated
with the primary vertex. Beam related background is suppressed
by rejecting events for which less than 25\% of all reconstructed tracks
pass the track selection criteria of this analysis.
The \pPb instantaneous luminosity provided by the LHC in the 2013
run resulted in an approximately 3\% probability of at least one additional interaction
occurring in the same bunch crossing.
Following the procedure developed in Ref.~\cite{Chatrchyan:2013nka} for rejecting such ``pileup'' events,
a 99.8\% purity of single-interaction events is achieved for the \pPb collisions
belonging to the highest multiplicity class studied in this Letter.
In \pPb interactions simulated with the \EPOS~\cite{Porteboeuf:2010um}
and \HIJING~\cite{Gyulassy:1994ew} event generators, requiring at least one primary
particle with total energy $E>3$\GeV in each of the $\eta$ ranges $-5<\eta<-3$ and $3<\eta<5$
is found to select 97--98\% of the total inelastic hadronic cross section.

The CMS ``high-quality" tracks, described in Ref.~\cite{Chatrchyan:2014fea}
are used in this analysis. Additionally, a reconstructed track is only considered as a
candidate track from the primary vertex if the significance of the separation along the beam axis ($z$)
between the track and the best vertex, $d_z/\sigma(d_z)$, and the significance
of the track-vertex impact parameter measured transverse to the beam,
$d_\mathrm{T}/\sigma(d_\mathrm{T})$, are each less than 3. The relative uncertainty
in the transverse-momentum measurement, $\sigma(\pt)/\pt$, is required to be less than 10\%.
To ensure high tracking efficiency and to reduce the rate of incorrectly reconstructed tracks,
only tracks within $\abs{\eta}<2.4$ and with $0.3 < \pt < 3.0$ \GeVc are used in the analysis.
A different \pt cutoff of 0.4\GeVc is used in the multiplicity determination because of constraints
on the online processing time for the HLT.

The entire \pPb data set is divided into classes of reconstructed track multiplicity, \noff.
The multiplicity classification in this analysis is identical to that used in Ref.~\cite{Chatrchyan:2013nka},
where more details are provided including a table relating \noff\ to the fraction of MB triggered events.
A subset of semi-peripheral \PbPb data collected during the
2011 LHC heavy-ion run with an MB trigger is also reanalyzed in order
to directly compare the \pPb and \PbPb systems at the same track multiplicity.
This \PbPb sample is reprocessed using the same event selection and track reconstruction
as for the present \pPb analysis.
A description of the 2011 \PbPb data can be found in Ref.~\cite{Chatrchyan:2012xq}.

Extending the previous two- and four-particle
azimuthal correlation measurements of Ref.~\cite{Chatrchyan:2013nka}, six- and eight-particle
azimuthal correlations~\cite{Bilandzic:2010jr} are evaluated in this analysis as:
\begin{equation}\begin{split}
\dmean{6} &\equiv
    \bigl<\!\bigl< \re^{in(\phi_{1} + \phi_{2} + \phi_{3} - \phi_{4} - \phi_{5}
    - \phi_{6})} \bigr>\!\bigr>,
\\
\dmean{8} &\equiv
    \bigl<\!\bigl< \re^{in(\phi_{1} + \phi_{2} + \phi_{3} + \phi_{4} - \phi_{5} - \phi_{6} - \phi_{7}
    - \phi_{8})} \bigr>\!\bigr>. \label{eq:corr}
\end{split}\end{equation}
Here $\phi_{i}$ $(i=1,...,8)$ are the
azimuthal angles of one unique combination of multiple particles in an event, $n$ is the harmonic number,
and $\bigl\langle\bigl\langle \cdots \bigr\rangle\bigr\rangle$ represents
the average over all combinations from all events within a given multiplicity range.
The corresponding cumulants, $\cn{6}$ and $\cn{8}$, are calculated as follows:
\ifthenelse{\boolean{cms@external}}{
\begin{equation}\begin{split}
\cn{6} =& \dmean{6} - 9 \cdot \dmean{4}\dmean{2} + 12 \cdot \dmean{2}^3,
\\
\cn{8} =& \dmean{8} - 16 \cdot \dmean{6}\dmean{2} -18 \cdot \dmean{4}^2 +
\\ &144 \cdot \dmean{4}\dmean{2}^2 - 144 \dmean{2}^4 , \label{eq:cn}
\end{split}\end{equation}
}{
\begin{equation}\begin{split}
\cn{6} =& \dmean{6} - 9 \cdot \dmean{4}\dmean{2} + 12 \cdot \dmean{2}^3,
\\
\cn{8} =& \dmean{8} - 16 \cdot \dmean{6}\dmean{2} -18 \cdot \dmean{4}^2 +
144 \cdot \dmean{4}\dmean{2}^2 - 144 \dmean{2}^4 , \label{eq:cn}
\end{split}\end{equation}
}
using the Q-cumulant method as formulated in Ref.~\cite{Bilandzic:2010jr}, where $\dmean{2}$ and $\dmean{4}$ are defined similarly as in Eq.~(\ref{eq:corr}).
The Fourier harmonics $v_n$ that characterize the global azimuthal behavior
are related to the multi-particle correlations~\cite{Bilandzic:2013kga} using
\begin{equation}\begin{split}
\vn{6} &= \sqrt[6]{\frac{1}{4} \cn{6}},
\\
\vn{8} &= \sqrt[8]{-\frac{1}{33} \cn{8}}. \label{eq:vn}
\end{split}\end{equation}

To account for detector effects, such as the tracking efficiency,
the Q-cumulant method was extended in Ref.~\cite{Bilandzic:2013kga}
to allow for particles having different weights.
Each reconstructed track is weighted by a correction factor to account
for the reconstruction efficiency, detector acceptance, and
fraction of misreconstructed tracks. This factor is derived  as a
function of \pt and $\eta$, as described in Refs.~\cite{Chatrchyan:2011eka,Chatrchyan:2012wg},
based on MC simulations. The combined geometrical acceptance and efficiency for track
reconstruction exceeds 60\% for $\pt \approx 0.3$\GeVc and $\abs{\eta}<2.4$.
The efficiency is greater than 90\% in the $\abs{\eta}<1$ region for $\pt>$ 0.6\GeVc.
For the entire multiplicity range (up to \noff\ $\sim$ 350) studied in this Letter,
no dependence of the tracking efficiency on multiplicity is found and the rate
of mis-reconstructed tracks remains at the 1--2\% level. The software package provided
by Ref.~\cite{Bilandzic:2013kga} is used to implement the weights of individual tracks
in the cumulant calculations.

The LYZ method~\cite{NuclPhysA.727.373, Borghini:2004ke} allows a direct study of the
large-order behavior by using the asymptotic form of the cumulant expansion to relate
locations of the zeros of a generating function to the azimuthal correlations.
This method has been employed in previous CMS PbPb analyses~\cite{Chatrchyan:2012ta,Chatrchyan:2013kba}.
For each multiplicity bin, the $v_2$ harmonic averaged over
$0.3 < p_T < 3.0 \GeVc$ is found using an integral generating function~\cite{Chatrchyan:2012ta}.
Similar to the cumulant methods, a weight
for each track is implemented to account for detector-related effects.  
In both methods, the statistical uncertainties are evaluated from data 
by dividing the data set into 20 subsets with roughly equal numbers of events 
and evaluating the standard deviation of the resulting distributions of the cumulant or $\vtwo{\text{LYZ}}$ values. 
In the case of low multiplicity or small flow signal, 
the LYZ method may overestimate the true collective flow. 
This effect was studied using MC pseudo-experiments for the event multiplicities covered in this analysis 
and a small correction is applied to the data. 
The correction is less than 3\% in the lowest multiplicity bin and becomes much smaller in higher-multiplicity bins. 
This correction is also included in the quoted LYZ systematic uncertainties.

Systematic uncertainties are estimated by varying the track quality requirements,
by comparing the results using efficiency correction tables from different MC event generators, and by
exploring the sensitivity of the results to the vertex position and to the \noff bin width.
For the \pPb data, potential HLT bias and pileup effects are also studied by
requiring the presence of only a single reconstructed vertex.
No evident \noff or beam direction dependent systematic effects are observed.
For \pPb collisions, a 5\% systematic uncertainty is obtained for $\vtwo{6}$ and a 6\%
uncertainty is found for both $\vtwo{8}$ and $\vtwo{\mathrm{LYZ}}$.
The corresponding uncertainties for \PbPb collisions are 2\% for $\vtwo{6}$ and $\vtwo{8}$, and  4\% for $\vtwo{\mathrm{LYZ}}$.

\begin{figure}[htb]
	\centering
   \includegraphics[width=\cmsFigWidth]{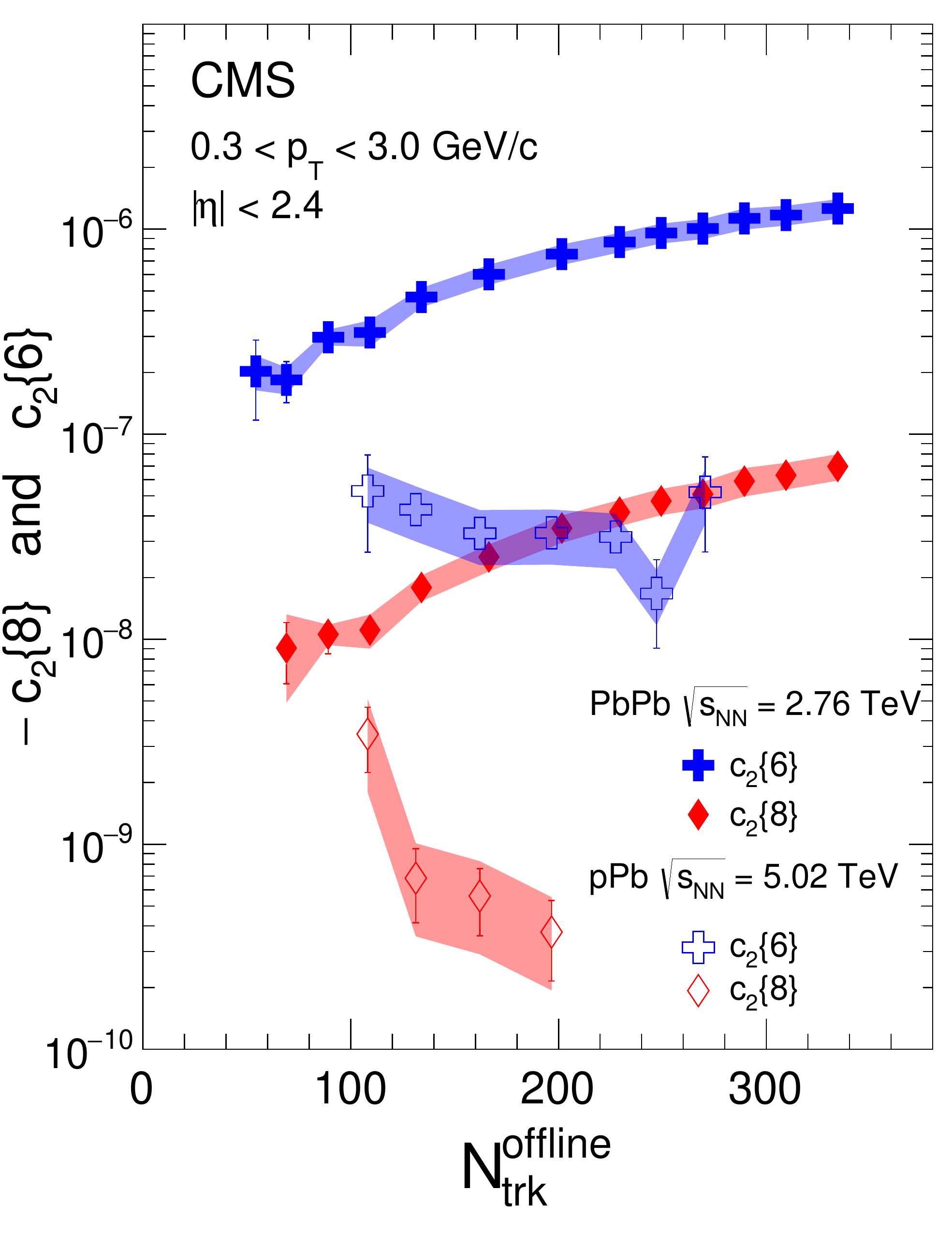}
	\caption{The cumulant $c_2\{6\}$ and $-c_2\{8\}$ results as a function of \noff
	         for \PbPb and \pPb reactions.
	         Error bars and shaded areas denote statistical and systematic uncertainties, respectively.}
	\label{fig:c2_cumu}
\end{figure}

In Fig.~\ref{fig:c2_cumu}, the six- and eight-particle cumulants, $\ctwo{6}$ and $\ctwo{8}$,
for particle \pt of 0.3--3.0\GeVc in 2.76\TeV PbPb and 5.02\TeV \pPb collisions
are shown as a function of event multiplicity.
The cumulants shown are required to be at least two standard deviations away from their physics boundaries ($\ctwo{6}/\sigma_{\ctwo{6}} > 2$, $\ctwo{8}/\sigma_{\ctwo{8}} < -2$),
so that the statistical uncertainties can be propagated as Gaussian fluctuations~\cite{PhysRevD.57.3873}.
Non-zero multi-particle correlation signals are observed in both \PbPb and \pPb collisions. The \pPb data exhibit larger statistical uncertainties than the \PbPb results,
mainly because of the smaller magnitudes of the correlation signals.
Because of the limited sample size,
the $\ctwo{6}$ and $\ctwo{8}$ values in \pPb collisions are derived for a smaller range in \noff.

\begin{figure}[htb]
	\centering
	\includegraphics[width=\columnwidth]{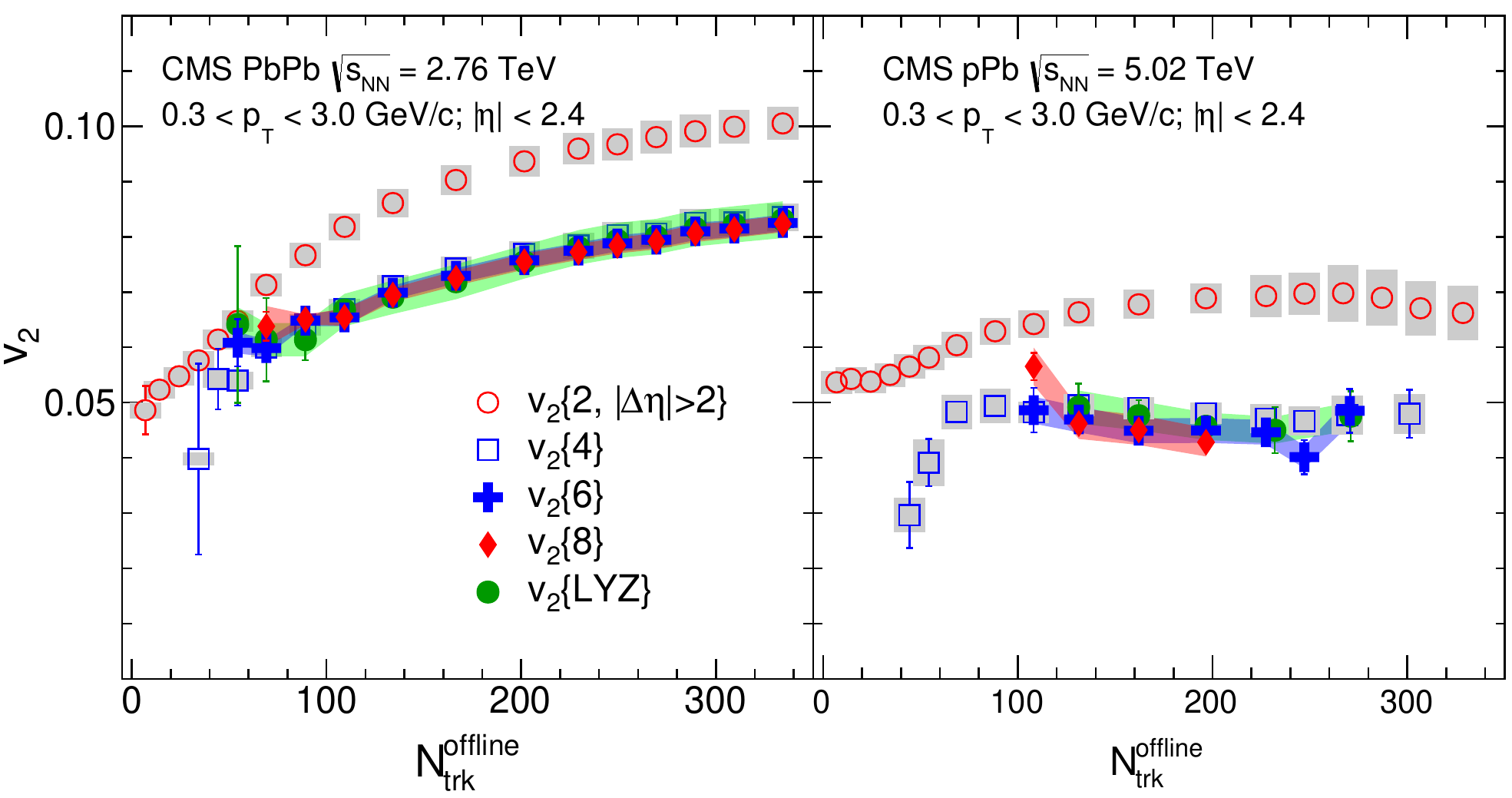}
	\caption{
	The $v_2$ values as a function of \noff. Open data points are published two- and four-particle $v_2$ results~\cite{Chatrchyan:2013nka}.
	Solid data points are $v_2$ results obtained from six- and eight-particle cumulants,
	and LYZ methods, averaged over the particle \pt range of 0.3--3.0\GeVc,
	in \PbPb at $\rootsNN = 2.76$\TeV (left) and \pPb at $\rootsNN = 5.02$\TeV (right).
	Statistical and systematic uncertainties are indicated by the error bars and shaded regions, respectively.}
	\label{fig:v2_cumu}
\end{figure}

The second-order anisotropy Fourier
harmonics, $v_2$, averaged over the \pt range of 0.3--3.0\GeVc,
are shown in Fig.~\ref{fig:v2_cumu} based on six- and eight-particle
cumulants (Eq.~(\ref{eq:vn})) for 2.76\TeV \PbPb (left) and 5.02\TeV \pPb (right)
collisions, as a function of event multiplicity. The open symbols are $v_2$
results extracted by CMS using two- and four-particle correlations~\cite{Chatrchyan:2013nka}.
The $v_2$ values derived using the LYZ method involving correlations among all particles
are also shown.
For each multiplicity bin, the values of $v_2\{4\}$, $v_2\{6\}$, $v_2\{8\}$, and $v_2\{\mathrm{LYZ}\}$ for \pPb collisions are found to be
in agreement within 10\%. 
For part of the multiplicity range, the values for $v_2\{4\}$ are larger than the others by a statistically significant amount, although still within 10\%.
The corresponding \PbPb values are consistently higher than for \pPb collisions, but within the \PbPb system
are found to be in agreement within 2\% for most multiplicity ranges and within 10\% for all multiplicities.
This supports the collective nature of the observed correlations, \ie, involving all particles from each system, and is inconsistent with a jet-related origin involving correlations among only a few particles.
The $v_2$ data from two-particle correlations are consistently above the multi-particle correlation data.
This behavior can be understood in hydrodynamic models,
where event-by-event participant geometry fluctuations of the $v_2$ coefficient
are expected to affect the two- and multi-particle cumulants differently~\cite{Ollitrault:2009ie,Ollitrault:2009it}.
Note that, to minimize jet-related non-flow effects, the $v_2\{2\}$ values are obtained with an $\eta$ gap of 2 units between the two particles.
Possible residual non-flow effects resulting from back-to-back jet correlations are
estimated using very low multiplicity events in Ref.~\cite{Chatrchyan:2013nka}.
Based on this analysis, such non-flow effects are expected to make a negligible contribution to $v_2\{2\}$ in very high multiplicity events.
In \PbPb collisions, the $v_2$ values from all methods show an increase with multiplicity, while little multiplicity dependence is
seen for the \pPb data.
This difference might reflect the presence of a lenticular overlap geometry in \PbPb collisions, which is not expected in \pPb collisions, that gives rise to a large (and varying) initial elliptic asymmetry in the PbPb system.

\begin{figure}[htb]
	\centering
	\includegraphics[width=0.49\textwidth]{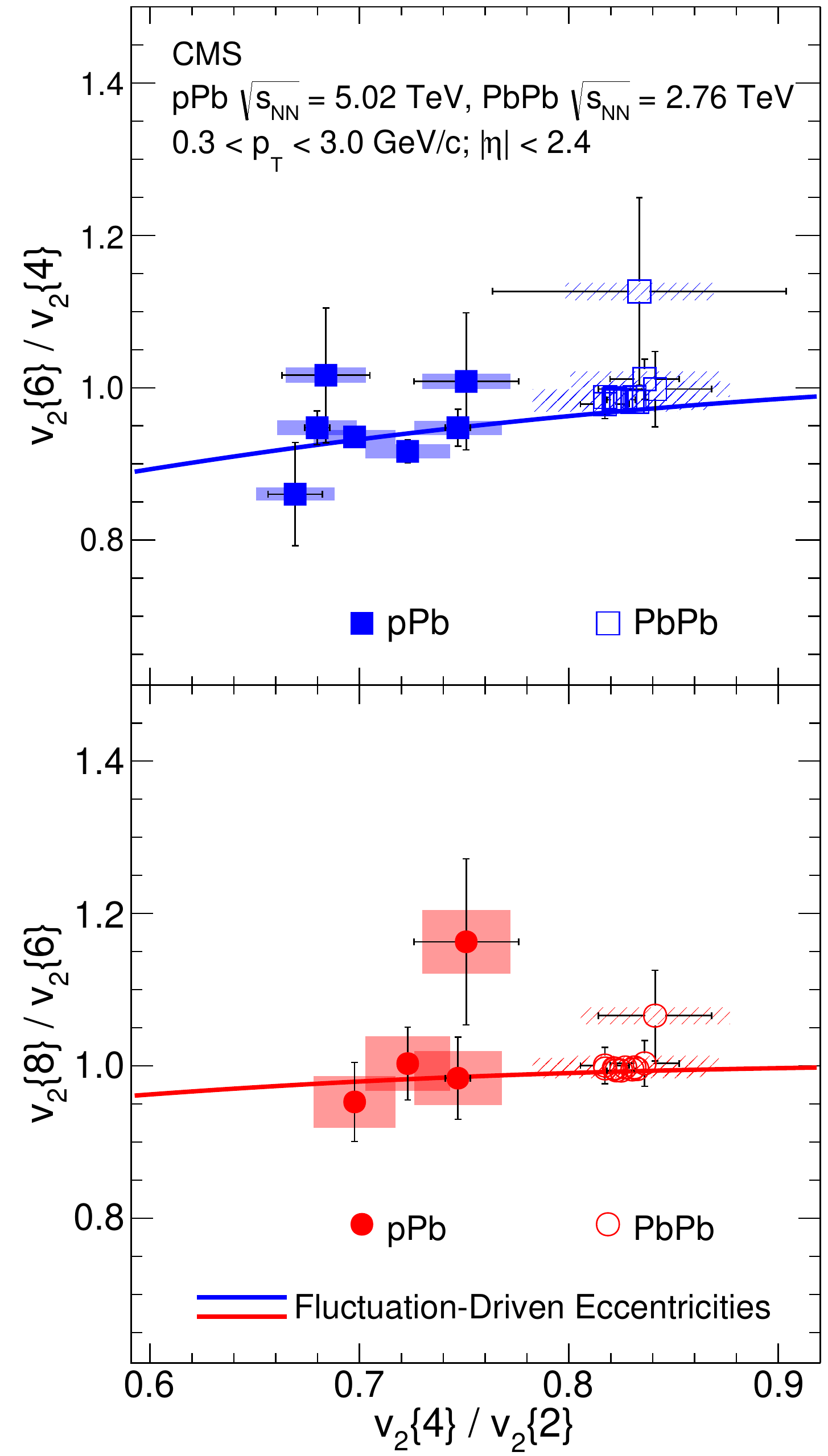}
	\caption{Cumulant ratios $v_2\{6\}/v_2\{4\}$ (top) and $v_2\{8\}/v_2\{6\}$ (bottom) as a function of
    	$v_2\{4\}/v_2\{2\}$ in \pPb collisions at $\rootsNN = 5.02$\TeV and \PbPb collisions at \rootsNN\ = 2.76\TeV. Error bars and shaded areas denote statistical and systematic uncertainties, respectively. The solid curves
    	show the expected behavior based on a hydrodynamics motivated study of the role of initial-state fluctuations~\cite{Yan:2013laa}.}
	\label{fig:pPb_v2_ratio}
\end{figure}

The effect of fluctuation-driven initial-state eccentricities on multi-particle cumulants has recently
been explored in the context of hydrodynamic behavior of the resulting medium~\cite{Yan:2013laa,Bzdak:2013rya}. For fluctuation-driven initial-state
conditions, ratios of $v_2$ values derived from various orders of multi-particle cumulants
are predicted to follow a universal behavior~\cite{Yan:2013laa}. In Fig.~\ref{fig:pPb_v2_ratio}, ratios of $v_2\{6\}/v_2\{4\}$ (top) and $v_2\{8\}/v_2\{6\}$ (bottom)
are calculated and plotted against $v_2\{4\}/v_2\{2\}$ in \pPb collisions at $\rootsNN = 5.02$\TeV.
The $v_2\{2\}$ and $v_2\{4\}$ data are taken from previously published CMS results~\cite{Chatrchyan:2013nka}.
The solid curves correspond to theoretical predictions for both large and small systems based on hydrodynamics and the assumption that the initial-state
geometry is purely driven by fluctuations~\cite{Yan:2013laa}.
The ratios from \PbPb collisions are also shown for comparison.
Note that the geometry of very central \PbPb collisions might be dominated by fluctuations,
but for these semi-peripheral \PbPb collisions the lenticular shape of the overlap region should also strongly contribute to the $v_2$ values.
The CMS \pPb data are consistent with the predictions
within statistical and systematic uncertainties.
The systematic uncertainties in the ratios presented in Fig.~\ref{fig:pPb_v2_ratio} are estimated to be 2.4\% for $\vtwo{4}/\vtwo{2}$ for both \pPb and \PbPb collisions,
1\% for $\vtwo{6}/\vtwo{4}$ in \pPb and \PbPb collisions, and 3.6\% and 1\% for $\vtwo{8}/\vtwo{6}$ in \pPb and \PbPb collisions, respectively.
Since they are all derived from the same data, the systematic uncertainties for the different cumulant orders are highly correlated and therefore partially cancel in the ratios.

Recently, other theoretical models based on quantum chromodynamics, and not involving hydrodynamics, have also been suggested to
explain the observed multi-particle correlations in \pPb collisions~\cite{Gyulassy:2014cfa,McLerran:2014uka}.
Unlike the descriptions based on hydrodynamic behavior, these models do not
require significant final-state interactions among quarks and gluons.
They suggest similar values for $v_2\{4\}$, $v_2\{6\}$, $v_2\{8\}$, and $v_2\{\mathrm{LYZ}\}$, without yet, however, providing quantitative predictions.

In summary, multi-particle azimuthal correlations among six, eight, and all particles
have been measured in \pPb collisions at $\rootsNN = 5.02$\TeV
by the CMS experiment. The new measurements extend previous CMS two- and four-particle
correlation analyses of \pPb collisions and strongly constrain possible explanations for the observed correlations.
A direct comparison of the correlation data for \pPb and \PbPb collisions
is presented as a function of particle multiplicity.
Averaging over the particle \pt range of 0.3--3.0\GeVc, multi-particle
correlation signals are observed in both \pPb and \PbPb collisions.
The second-order azimuthal anisotropy
Fourier harmonic, $v_2$, is extracted using six- and eight-particle cumulants
and using the LYZ method which involves all particles.
The $v_2$ values obtained using correlation methods including four or more particles
are consistent within $\pm$2\% for the \PbPb system, and within $\pm$10\% for the \pPb system.
This measurement supports the collective nature of the observed correlations.
The ratios of $v_2$ values obtained using different numbers of particles are found to be consistent with
hydrodynamic model calculations for \pPb collisions.

\begin{acknowledgments}
We congratulate our colleagues in the CERN accelerator departments for the excellent performance of the LHC and thank the technical and administrative staffs at CERN and at other CMS institutes for their contributions to the success of the CMS effort. In addition, we gratefully acknowledge the computing centers and personnel of the Worldwide LHC Computing Grid for delivering so effectively the computing infrastructure essential to our analyses. Finally, we acknowledge the enduring support for the construction and operation of the LHC and the CMS detector provided by the following funding agencies: BMWFW and FWF (Austria); FNRS and FWO (Belgium); CNPq, CAPES, FAPERJ, and FAPESP (Brazil); MES (Bulgaria); CERN; CAS, MoST, and NSFC (China); COLCIENCIAS (Colombia); MSES and CSF (Croatia); RPF (Cyprus); MoER, ERC IUT and ERDF (Estonia); Academy of Finland, MEC, and HIP (Finland); CEA and CNRS/IN2P3 (France); BMBF, DFG, and HGF (Germany); GSRT (Greece); OTKA and NIH (Hungary); DAE and DST (India); IPM (Iran); SFI (Ireland); INFN (Italy); MSIP and NRF (Republic of Korea); LAS (Lithuania); MOE and UM (Malaysia); CINVESTAV, CONACYT, SEP, and UASLP-FAI (Mexico); MBIE (New Zealand); PAEC (Pakistan); MSHE and NSC (Poland); FCT (Portugal); JINR (Dubna); MON, RosAtom, RAS and RFBR (Russia); MESTD (Serbia); SEIDI and CPAN (Spain); Swiss Funding Agencies (Switzerland); MST (Taipei); ThEPCenter, IPST, STAR and NSTDA (Thailand); TUBITAK and TAEK (Turkey); NASU and SFFR (Ukraine); STFC (United Kingdom); DOE and NSF (USA). \end{acknowledgments}

\bibliography{auto_generated}

\providecommand{\href}[2]{#2}\begingroup\raggedright\begin{thebibliography}{10}%
\makeatletter
\providecommand{\hrefCMSnoop }[0]{\@secondoftwo}%
\makeatother
\providecommand{\doi}{\texttt{doi:}\begingroup \urlstyle{tt}\Url}

\bibitem{Khachatryan:2010gv}
\hrefCMSnoop {}{{CMS} Collaboration, ``{Observation of Long-Range Near-Side
  Angular Correlations in Proton-Proton Collisions at the LHC}'',} \textit{
  JHEP} \textbf{ 09} (2010) 091,
  \href{http://dx.doi.org/10.1007/JHEP09(2010)091}{\doi{10.1007/JHEP09(2010)091}},
\href{http://www.arXiv.org/abs/1009.4122}{\texttt{arXiv:1009.4122}}.

\bibitem{CMS:2012qk}
\hrefCMSnoop {}{{CMS} Collaboration, ``{Observation of long-range near-side
  angular correlations in proton-lead collisions at the LHC}'',} \textit{ Phys.
  Lett. B} \textbf{ 718} (2013) 795,
  \href{http://dx.doi.org/10.1016/j.physletb.2012.11.025}{\doi{10.1016/j.physletb.2012.11.025}},
\href{http://www.arXiv.org/abs/1210.5482}{\texttt{arXiv:1210.5482}}.

\bibitem{alice:2012qe}
\hrefCMSnoop {}{{ALICE} Collaboration, ``{Long-range angular correlations on
  the near and away side in \pPb\ collisions at \rootsNN\ = 5.02\TeV }'',}
  \textit{ Phys. Lett. B} \textbf{ 719} (2013) 29,
  \href{http://dx.doi.org/10.1016/j.physletb.2013.01.012}{\doi{10.1016/j.physletb.2013.01.012}},
\href{http://www.arXiv.org/abs/1212.2001}{\texttt{arXiv:1212.2001}}.

\bibitem{atlas:2012fa}
\hrefCMSnoop {}{{ATLAS} Collaboration, ``{Observation of Associated Near-side
  and Away-side Long-range Correlations in \rootsNN\ = 5.02 TeV Proton-lead
  Collisions with the ATLAS Detector}'',} \textit{ Phys. Rev. Lett.} \textbf{
  110} (2013) 182302,
  \href{http://dx.doi.org/10.1103/PhysRevLett.110.182302}{\doi{10.1103/PhysRevLett.110.182302}},
\href{http://www.arXiv.org/abs/1212.5198}{\texttt{arXiv:1212.5198}}.

\bibitem{Abelev:2014mda}
\hrefCMSnoop {}{{ALICE} Collaboration, ``{Multiparticle azimuthal correlations
  in p-Pb and Pb-Pb collisions at the CERN Large Hadron Collider}'',} \textit{
  Phys. Rev. C} \textbf{ 90} (2014) 054901,
  \href{http://dx.doi.org/10.1103/PhysRevC.90.054901}{\doi{10.1103/PhysRevC.90.054901}},
\href{http://www.arXiv.org/abs/1406.2474}{\texttt{arXiv:1406.2474}}.

\bibitem{Adare:2013piz}
\hrefCMSnoop {}{{PHENIX} Collaboration, ``{Quadrupole Anisotropy in Dihadron
  Azimuthal Correlations in Central $d$$+$Au Collisions at \rootsNN\ = 200
  GeV}'',} \textit{ Phys. Rev. Lett.} \textbf{ 111} (2013) 212301,
  \href{http://dx.doi.org/10.1103/PhysRevLett.111.212301}{\doi{10.1103/PhysRevLett.111.212301}},
\href{http://www.arXiv.org/abs/1303.1794}{\texttt{arXiv:1303.1794}}.

\bibitem{Adare:2014keg}
\hrefCMSnoop {}{{PHENIX} Collaboration, ``{Measurement of long-range angular
  correlation and quadrupole anisotropy of pions and (anti)protons in central
  $d$$+$Au collisions at $\sqrt{s_{_{NN}}}$=200 GeV}'',} \textit{ Phys. Rev.
  Lett.} \textbf{ 114} (2015) 192301,
  \href{http://dx.doi.org/10.1103/PhysRevLett.114.192301}{\doi{10.1103/PhysRevLett.114.192301}},
\href{http://www.arXiv.org/abs/1404.7461}{\texttt{arXiv:1404.7461}}.

\bibitem{Alver:2008gk}
\hrefCMSnoop {}{{PHOBOS} Collaboration, ``{System size dependence of cluster
  properties from two- particle angular correlations in Cu+Cu and Au+Au
  collisions at \rootsNN\ = 200 GeV}'',} \textit{ Phys. Rev. C} \textbf{ 81}
  (2010) 024904,
  \href{http://dx.doi.org/10.1103/PhysRevC.81.024904}{\doi{10.1103/PhysRevC.81.024904}},
\href{http://www.arXiv.org/abs/0812.1172}{\texttt{arXiv:0812.1172}}.

\bibitem{Adams:2005ph}
\hrefCMSnoop {}{{STAR} Collaboration, ``{Distributions of charged hadrons
  associated with high transverse momentum particles in pp and Au + Au
  collisions at \rootsNN\ = 200\GeV}'',} \textit{ Phys. Rev. Lett.} \textbf{
  95} (2005) 152301,
  \href{http://dx.doi.org/10.1103/PhysRevLett.95.152301}{\doi{10.1103/PhysRevLett.95.152301}},
\href{http://www.arXiv.org/abs/nucl-ex/0501016}{\texttt{arXiv:nucl-ex/0501016}}.

\bibitem{Abelev:2009af}
\hrefCMSnoop {}{{STAR} Collaboration, ``{Long range rapidity correlations and
  jet production in high energy nuclear collisions}'',} \textit{ Phys. Rev. C}
  \textbf{ 80} (2009) 064912,
  \href{http://dx.doi.org/10.1103/PhysRevC.80.064912}{\doi{10.1103/PhysRevC.80.064912}},
\href{http://www.arXiv.org/abs/0909.0191}{\texttt{arXiv:0909.0191}}.

\bibitem{Alver:2009id}
\hrefCMSnoop {}{{PHOBOS} Collaboration, ``{High transverse momentum triggered
  correlations over a large pseudorapidity acceptance in Au+Au collisions at
  \rootsNN\ = 200 GeV}'',} \textit{ Phys. Rev. Lett.} \textbf{ 104} (2010)
  062301,
  \href{http://dx.doi.org/10.1103/PhysRevLett.104.062301}{\doi{10.1103/PhysRevLett.104.062301}},
\href{http://www.arXiv.org/abs/0903.2811}{\texttt{arXiv:0903.2811}}.

\bibitem{Abelev:2009jv}
\hrefCMSnoop {}{{STAR} Collaboration, ``{Three-particle coincidence of the long
  range pseudorapidity correlation in high energy nucleus-nucleus
  collisions}'',} \textit{ Phys. Rev. Lett.} \textbf{ 105} (2010) 022301,
  \href{http://dx.doi.org/10.1103/PhysRevLett.105.022301}{\doi{10.1103/PhysRevLett.105.022301}},
\href{http://www.arXiv.org/abs/0912.3977}{\texttt{arXiv:0912.3977}}.

\bibitem{Chatrchyan:2011eka}
\hrefCMSnoop {}{{CMS} Collaboration, ``{Long-range and short-range dihadron
  angular correlations in central PbPb collisions at a nucleon-nucleon center
  of mass energy of 2.76 TeV}'',} \textit{ JHEP} \textbf{ 07} (2011) 076,
  \href{http://dx.doi.org/10.1007/JHEP07(2011)076}{\doi{10.1007/JHEP07(2011)076}},
\href{http://www.arXiv.org/abs/1105.2438}{\texttt{arXiv:1105.2438}}.

\bibitem{Chatrchyan:2012wg}
\hrefCMSnoop {}{{CMS} Collaboration, ``{Centrality dependence of dihadron
  correlations and azimuthal anisotropy harmonics in PbPb collisions at
  \rootsNN\ = 2.76 TeV}'',} \textit{ Eur. Phys. J. C} \textbf{ 72} (2012) 2012,
  \href{http://dx.doi.org/10.1140/epjc/s10052-012-2012-3}{\doi{10.1140/epjc/s10052-012-2012-3}},
\href{http://www.arXiv.org/abs/1201.3158}{\texttt{arXiv:1201.3158}}.

\bibitem{Aamodt:2011by}
\hrefCMSnoop {}{{ALICE} Collaboration, ``{Harmonic decomposition of
  two-particle angular correlations in Pb-Pb collisions at \rootsNN\ =
  2.76\TeV}'',} \textit{ Phys. Lett. B} \textbf{ 708} (2012) 249,
  \href{http://dx.doi.org/10.1016/j.physletb.2012.01.060}{\doi{10.1016/j.physletb.2012.01.060}},
\href{http://www.arXiv.org/abs/1109.2501}{\texttt{arXiv:1109.2501}}.

\bibitem{ATLAS:2012at}
\hrefCMSnoop {}{{ATLAS} Collaboration, ``{Measurement of the azimuthal
  anisotropy for charged particle production in \rootsNN\ = 2.76\TeV lead-lead
  collisions with the ATLAS detector}'',} \textit{ Phys. Rev. C} \textbf{ 86}
  (2012) 014907,
  \href{http://dx.doi.org/10.1103/PhysRevC.86.014907}{\doi{10.1103/PhysRevC.86.014907}},
\href{http://www.arXiv.org/abs/1203.3087}{\texttt{arXiv:1203.3087}}.

\bibitem{Chatrchyan:2012ta}
\hrefCMSnoop {}{{CMS} Collaboration, ``{Measurement of the elliptic anisotropy
  of charged particles produced in PbPb collisions at $\sqrt{s_{_{NN}}}$= 2.76
  TeV}'',} \textit{ Phys. Rev. C} \textbf{ 87} (2013) 014902,
  \href{http://dx.doi.org/10.1103/PhysRevC.87.014902}{\doi{10.1103/PhysRevC.87.014902}},
\href{http://www.arXiv.org/abs/1204.1409}{\texttt{arXiv:1204.1409}}.

\bibitem{CMS:2013bza}
\hrefCMSnoop {}{{CMS} Collaboration, ``{Studies of azimuthal dihadron
  correlations in ultra-central PbPb collisions at \rootsNN\ = 2.76\TeV}'',}
  \textit{ JHEP} \textbf{ 02} (2014) 088,
  \href{http://dx.doi.org/10.1007/JHEP02(2014)088}{\doi{10.1007/JHEP02(2014)088}},
\href{http://www.arXiv.org/abs/1312.1845}{\texttt{arXiv:1312.1845}}.

\bibitem{Ollitrault:1992bk}
\hrefCMSnoop {}{J.-Y. Ollitrault, ``{Anisotropy as a signature of transverse
  collective flow}'',} \textit{ Phys. Rev. D} \textbf{ 46} (1992) 229,
\href{http://dx.doi.org/10.1103/PhysRevD.46.229}{\doi{10.1103/PhysRevD.46.229}}.

\bibitem{Heinz:2013th}
\hrefCMSnoop {}{U.~Heinz and R.~Snellings, ``{Collective flow and viscosity in
  relativistic heavy-ion collisions}'',} \textit{ Ann. Rev. Nucl. Part. Sci.}
  \textbf{ 63} (2013) 123,
  \href{http://dx.doi.org/10.1146/annurev-nucl-102212-170540}{\doi{10.1146/annurev-nucl-102212-170540}},
\href{http://www.arXiv.org/abs/1301.2826}{\texttt{arXiv:1301.2826}}.

\bibitem{Gale:2013da}
\hrefCMSnoop {}{C.~Gale, S.~Jeon, and B.~Schenke, ``{Hydrodynamic Modeling of
  Heavy-Ion Collisions}'',} \textit{ Int. J. Mod. Phys. A} \textbf{ 28} (2013)
  1340011,
  \href{http://dx.doi.org/10.1142/S0217751X13400113}{\doi{10.1142/S0217751X13400113}},
\href{http://www.arXiv.org/abs/1301.5893}{\texttt{arXiv:1301.5893}}.

\bibitem{Li:2012hc}
\hrefCMSnoop {}{W.~Li, ``{Observation of a 'Ridge' correlation structure in
  high multiplicity proton-proton collisions: A brief review}'',} \textit{ Mod.
  Phys. Lett. A} \textbf{ 27} (2012) 1230018,
  \href{http://dx.doi.org/10.1142/S0217732312300182}{\doi{10.1142/S0217732312300182}},
\href{http://www.arXiv.org/abs/1206.0148}{\texttt{arXiv:1206.0148}}.

\bibitem{Bjorken:2013boa}
\hrefCMSnoop {}{J.~D. Bjorken, S.~J. Brodsky, and A.~Scharff~Goldhaber,
  ``{Possible multiparticle ridge-like correlations in very high multiplicity
  proton-proton collisions}'',} \textit{ Phys. Lett. B} \textbf{ 726} (2013)
  344,
  \href{http://dx.doi.org/10.1016/j.physletb.2013.08.066}{\doi{10.1016/j.physletb.2013.08.066}},
\href{http://www.arXiv.org/abs/1308.1435}{\texttt{arXiv:1308.1435}}.

\bibitem{Dusling:2012wy}
\hrefCMSnoop {}{K.~Dusling and R.~Venugopalan, ``{Explanation of systematics of
  CMS p+Pb high multiplicity di-hadron data at \rootsNN\ = 5.02 TeV}'',}
  \textit{ Phys. Rev. D} \textbf{ 87} (2013) 054014,
  \href{http://dx.doi.org/10.1103/PhysRevD.87.054014}{\doi{10.1103/PhysRevD.87.054014}},
\href{http://www.arXiv.org/abs/1211.3701}{\texttt{arXiv:1211.3701}}.

\bibitem{Dusling:2012cg}
\hrefCMSnoop {}{K.~Dusling and R.~Venugopalan, ``{Evidence for BFKL and
  saturation dynamics from dihadron spectra at the LHC}'',} \textit{ Phys. Rev.
  D} \textbf{ 87} (2013) 051502,
  \href{http://dx.doi.org/10.1103/PhysRevD.87.051502}{\doi{10.1103/PhysRevD.87.051502}},
\href{http://www.arXiv.org/abs/1210.3890}{\texttt{arXiv:1210.3890}}.

\bibitem{Schenke:2014zha}
\hrefCMSnoop {}{B.~Schenke and R.~Venugopalan, ``{Eccentric protons?
  Sensitivity of flow to system size and shape in p+p, p+Pb and Pb+Pb
  collisions}'',} \textit{ Phys. Rev. Lett.} \textbf{ 113} (2014) 102301,
  \href{http://dx.doi.org/10.1103/PhysRevLett.113.102301}{\doi{10.1103/PhysRevLett.113.102301}},
\href{http://www.arXiv.org/abs/1405.3605}{\texttt{arXiv:1405.3605}}.

\bibitem{Bozek:2011if}
\hrefCMSnoop {}{P.~Bozek, ``{Collective flow in p-Pb and d-Pd collisions at TeV
  energies}'',} \textit{ Phys. Rev. C} \textbf{ 85} (2012) 014911,
  \href{http://dx.doi.org/10.1103/PhysRevC.85.014911}{\doi{10.1103/PhysRevC.85.014911}},
\href{http://www.arXiv.org/abs/1112.0915}{\texttt{arXiv:1112.0915}}.

\bibitem{Bozek:2012gr}
\hrefCMSnoop {}{P.~Bozek and W.~Broniowski, ``{Correlations from hydrodynamic
  flow in \pPb\ collisions}'',} \textit{ Phys. Lett. B} \textbf{ 718} (2013)
  1557,
  \href{http://dx.doi.org/10.1016/j.physletb.2012.12.051}{\doi{10.1016/j.physletb.2012.12.051}},
\href{http://www.arXiv.org/abs/1211.0845}{\texttt{arXiv:1211.0845}}.

\bibitem{Voloshin:1994mz}
\hrefCMSnoop {}{S.~Voloshin and Y.~Zhang, ``{Flow study in relativistic nuclear
  collisions by Fourier expansion of azimuthal particle distributions}'',}
  \textit{ Z. Phys. C} \textbf{ 70} (1996) 665,
  \href{http://dx.doi.org/10.1007/s002880050141}{\doi{10.1007/s002880050141}},
\href{http://www.arXiv.org/abs/hep-ph/9407282}{\texttt{arXiv:hep-ph/9407282}}.

\bibitem{Alver:2010gr}
\hrefCMSnoop {}{B.~Alver and G.~Roland, ``{Collision geometry fluctuations and
  triangular flow in heavy-ion collisions}'',} \textit{ Phys. Rev. C} \textbf{
  81} (2010) 054905,
  \href{http://dx.doi.org/10.1103/PhysRevC.81.054905}{\doi{10.1103/PhysRevC.81.054905}},
  \href{http://www.arXiv.org/abs/1003.0194}{\texttt{arXiv:1003.0194}}.
[Erratum: \DOI{10.1103/PhysRevC.82.039903}].

\bibitem{Alver:2010dn}
\hrefCMSnoop {}{B.~H. Alver, C.~Gombeaud, M.~Luzum, and J.-Y. Ollitrault,
  ``{Triangular flow in hydrodynamics and transport theory}'',} \textit{ Phys.
  Rev. C} \textbf{ 82} (2010) 034913,
  \href{http://dx.doi.org/10.1103/PhysRevC.82.034913}{\doi{10.1103/PhysRevC.82.034913}},
\href{http://www.arXiv.org/abs/1007.5469}{\texttt{arXiv:1007.5469}}.

\bibitem{Schenke:2010rr}
\hrefCMSnoop {}{B.~Schenke, S.~Jeon, and C.~Gale, ``Elliptic and triangular
  flow in event-by-event {D=3+1} viscous hydrodynamics'',} \textit{ Phys. Rev.
  Lett.} \textbf{ 106} (2011) 042301,
  \href{http://dx.doi.org/10.1103/PhysRevLett.106.042301}{\doi{10.1103/PhysRevLett.106.042301}},
\href{http://www.arXiv.org/abs/1009.3244}{\texttt{arXiv:1009.3244}}.

\bibitem{Qiu:2011hf}
\hrefCMSnoop {}{Z.~Qiu, C.~Shen, and U.~Heinz, ``{Hydrodynamic elliptic and
  triangular flow in Pb-Pb collisions at \rootsNN\ = 2.76A TeV}'',} \textit{
  Phys. Lett. B} \textbf{ 707} (2012) 151,
  \href{http://dx.doi.org/10.1016/j.physletb.2011.12.041}{\doi{10.1016/j.physletb.2011.12.041}},
\href{http://www.arXiv.org/abs/1110.3033}{\texttt{arXiv:1110.3033}}.

\bibitem{Aad:2013fja}
\hrefCMSnoop {}{{ATLAS} Collaboration, ``{Measurement with the ATLAS detector
  of multi-particle azimuthal correlations in p+Pb collisions at \rootsNN\ =
  5.02 TeV}'',} \textit{ Phys. Lett. B} \textbf{ 725} (2013) 60,
  \href{http://dx.doi.org/10.1016/j.physletb.2013.06.057}{\doi{10.1016/j.physletb.2013.06.057}},
\href{http://www.arXiv.org/abs/1303.2084}{\texttt{arXiv:1303.2084}}.

\bibitem{Chatrchyan:2013nka}
\hrefCMSnoop {}{{CMS} Collaboration, ``{Multiplicity and transverse momentum
  dependence of two- and four-particle correlations in pPb and PbPb
  collisions}'',} \textit{ Phys. Lett. B} \textbf{ 724} (2013) 213,
  \href{http://dx.doi.org/10.1016/j.physletb.2013.06.028}{\doi{10.1016/j.physletb.2013.06.028}},
\href{http://www.arXiv.org/abs/1305.0609}{\texttt{arXiv:1305.0609}}.

\bibitem{Bilandzic:2010jr}
\hrefCMSnoop {}{A.~Bilandzic, R.~Snellings, and S.~Voloshin, ``{Flow analysis
  with cumulants: Direct calculations}'',} \textit{ Phys. Rev. C} \textbf{ 83}
  (2011) 044913,
  \href{http://dx.doi.org/10.1103/PhysRevC.83.044913}{\doi{10.1103/PhysRevC.83.044913}},
\href{http://www.arXiv.org/abs/1010.0233}{\texttt{arXiv:1010.0233}}.

\bibitem{NuclPhysA.727.373}
\hrefCMSnoop {}{R.~S. Bhalerao, N.~Borghini, and J.~Y. Ollitrault, ``Analysis
  of anisotropic flow with Lee-Yang zeroes'',} \textit{ Nucl. Phys. A} \textbf{
  727} (2003) 373,
  \href{http://dx.doi.org/10.1016/j.nuclphysa.2003.08.007}{\doi{10.1016/j.nuclphysa.2003.08.007}},
  \href{http://www.arXiv.org/abs/nucl-th/0310016}{\texttt{arXiv:nucl-th/0310016}}.

\bibitem{Borghini:2004ke}
\hrefCMSnoop {}{N.~Borghini, R.~S. Bhalerao, and J.~Y. Ollitrault,
  ``{Anisotropic flow from Lee-Yang zeroes: A practical guide}'',} \textit{ J.
  Phys. G} \textbf{ 30} (2004) S1213,
  \href{http://dx.doi.org/10.1088/0954-3899/30/8/092}{\doi{10.1088/0954-3899/30/8/092}},
\href{http://www.arXiv.org/abs/nucl-th/0402053}{\texttt{arXiv:nucl-th/0402053}}.

\bibitem{JINST}
\hrefCMSnoop {}{{CMS} Collaboration, ``The {CMS} experiment at the {CERN}
  {LHC}'',} \textit{ JINST} \textbf{ 3} (2008) S08004,
\href{http://dx.doi.org/10.1088/1748-0221/3/08/S08004}{\doi{10.1088/1748-0221/3/08/S08004}}.

\bibitem{GEANT4}
\hrefCMSnoop {}{{Geant4} Collaboration, ``{Geant4: A simulation toolkit}'',}
  \textit{ Nucl. Instrum. Meth. A} \textbf{ 506} (2003) 250,
\href{http://dx.doi.org/10.1016/S0168-9002(03)01368-8}{\doi{10.1016/S0168-9002(03)01368-8}}.

\bibitem{Porteboeuf:2010um}
\hrefCMSnoop {}{S.~Porteboeuf, T.~Pierog, and K.~Werner, ``{Producing Hard
  Processes Regarding the Complete Event: The EPOS Event Generator}'',} (2010).
\href{http://www.arXiv.org/abs/1006.2967}{\texttt{arXiv:1006.2967}}.

\bibitem{Gyulassy:1994ew}
\hrefCMSnoop {}{M.~Gyulassy and X.-N. Wang, ``{HIJING 1.0: A Monte Carlo
  program for parton and particle production in high-energy hadronic and
  nuclear collisions}'',} \textit{ Comput. Phys. Commun.} \textbf{ 83} (1994)
  307,
  \href{http://dx.doi.org/10.1016/0010-4655(94)90057-4}{\doi{10.1016/0010-4655(94)90057-4}},
\href{http://www.arXiv.org/abs/nucl-th/9502021}{\texttt{arXiv:nucl-th/9502021}}.

\bibitem{Chatrchyan:2014fea}
\hrefCMSnoop {}{{CMS} Collaboration, ``{Description and performance of track
  and primary-vertex reconstruction with the CMS tracker}'',} \textit{ JINST}
  \textbf{ 9} (2014) P10009,
  \href{http://dx.doi.org/10.1088/1748-0221/9/10/P10009}{\doi{10.1088/1748-0221/9/10/P10009}},
\href{http://www.arXiv.org/abs/1405.6569}{\texttt{arXiv:1405.6569}}.

\bibitem{Chatrchyan:2012xq}
\hrefCMSnoop {}{{CMS} Collaboration, ``{Azimuthal anisotropy of charged
  particles at high transverse momenta in PbPb collisions at \rootsNN\ =
  2.76\TeV}'',} \textit{ Phys. Rev. Lett.} \textbf{ 109} (2012) 022301,
  \href{http://dx.doi.org/10.1103/PhysRevLett.109.022301}{\doi{10.1103/PhysRevLett.109.022301}},
\href{http://www.arXiv.org/abs/1204.1850}{\texttt{arXiv:1204.1850}}.

\bibitem{Bilandzic:2013kga}
A.~Bilandzic\hrefCMSnoop {}{ {et~al.}, ``{Generic framework for anisotropic
  flow analyses with multi-particle azimuthal correlations}'',} \textit{ Phys.
  Rev. C} \textbf{ 89} (2014) 064904,
  \href{http://dx.doi.org/10.1103/PhysRevC.89.064904}{\doi{10.1103/PhysRevC.89.064904}},
\href{http://www.arXiv.org/abs/1312.3572}{\texttt{arXiv:1312.3572}}.

\bibitem{Chatrchyan:2013kba}
\hrefCMSnoop {}{{CMS} Collaboration, ``{Measurement of higher-order harmonic
  azimuthal anisotropy in PbPb collisions at \rootsNN\ = 2.76 TeV}'',} \textit{
  Phys. Rev. C} \textbf{ 89} (2014) 044906,
  \href{http://dx.doi.org/10.1103/PhysRevC.89.044906}{\doi{10.1103/PhysRevC.89.044906}},
\href{http://www.arXiv.org/abs/1310.8651}{\texttt{arXiv:1310.8651}}.

\bibitem{PhysRevD.57.3873}
\hrefCMSnoop {}{G.~J. Feldman and R.~D. Cousins, ``Unified approach to the
  classical statistical analysis of small signals'',} \textit{ Phys. Rev. D}
  \textbf{ 57} (1998) 3873,
  \href{http://dx.doi.org/10.1103/PhysRevD.57.3873}{\doi{10.1103/PhysRevD.57.3873}},
  \href{http://www.arXiv.org/abs/physics/9711021}{\texttt{arXiv:physics/9711021}}.

\bibitem{Ollitrault:2009ie}
\hrefCMSnoop {}{J.-Y. Ollitrault, A.~M. Poskanzer, and S.~A. Voloshin,
  ``{Effect of flow fluctuations and nonflow on elliptic flow methods}'',}
  \textit{ Phys. Rev. C} \textbf{ 80} (2009) 014904,
  \href{http://dx.doi.org/10.1103/PhysRevC.80.014904}{\doi{10.1103/PhysRevC.80.014904}},
\href{http://www.arXiv.org/abs/0904.2315}{\texttt{arXiv:0904.2315}}.

\bibitem{Ollitrault:2009it}
\hrefCMSnoop {}{J.-Y. Ollitrault, A.~M. Poskanzer, and S.~A. Voloshin,
  ``{Effect of nonflow and flow fluctuations on elliptic flow methods}'',}
  \textit{ Nucl. Phys. A} \textbf{ 830} (2009) 279C,
  \href{http://dx.doi.org/10.1016/j.nuclphysa.2009.09.026}{\doi{10.1016/j.nuclphysa.2009.09.026}},
\href{http://www.arXiv.org/abs/0906.3463}{\texttt{arXiv:0906.3463}}.

\bibitem{Yan:2013laa}
\hrefCMSnoop {}{L.~Yan and J.-Y. Ollitrault, ``{Universal fluctuation-driven
  eccentricities in proton-proton, proton-nucleus and nucleus-nucleus
  collisions}'',} \textit{ Phys. Rev. Lett.} \textbf{ 112} (2014) 082301,
  \href{http://dx.doi.org/10.1103/PhysRevLett.112.082301}{\doi{10.1103/PhysRevLett.112.082301}},
\href{http://www.arXiv.org/abs/1312.6555}{\texttt{arXiv:1312.6555}}.

\bibitem{Bzdak:2013rya}
\hrefCMSnoop {}{A.~Bzdak, P.~Bozek, and L.~McLerran, ``{Fluctuation induced
  equality of multi-particle eccentricities for four or more particles}'',}
  \textit{ Nucl. Phys. A} \textbf{ 927} (2014) 15,
  \href{http://dx.doi.org/10.1016/j.nuclphysa.2014.03.007}{\doi{10.1016/j.nuclphysa.2014.03.007}},
\href{http://www.arXiv.org/abs/1311.7325}{\texttt{arXiv:1311.7325}}.

\bibitem{Gyulassy:2014cfa}
\hrefCMSnoop {}{M.~Gyulassy, P.~Levai, I.~Vitev, and T.~S. Bir{\'o},
  ``{Non-Abelian Bremsstrahlung and Azimuthal Asymmetries in High Energy p+A
  Reactions}'',} \textit{ Phys. Rev. D} \textbf{ 90} (2014) 054025,
  \href{http://dx.doi.org/10.1103/PhysRevD.90.054025}{\doi{10.1103/PhysRevD.90.054025}},
\href{http://www.arXiv.org/abs/1405.7825}{\texttt{arXiv:1405.7825}}.

\bibitem{McLerran:2014uka}
\hrefCMSnoop {}{L.~McLerran and V.~V. Skokov, ``{The Eccentric Collective BFKL
  Pomeron}'',} (2014).
\href{http://www.arXiv.org/abs/1407.2651}{\texttt{arXiv:1407.2651}}.

\end{thebibliography}\endgroup
\cleardoublepage \appendix\section{The CMS Collaboration \label{app:collab}}\begin{sloppypar}\hyphenpenalty=5000\widowpenalty=500\clubpenalty=5000\textbf{Yerevan Physics Institute,  Yerevan,  Armenia}\\*[0pt]
V.~Khachatryan, A.M.~Sirunyan, A.~Tumasyan
\vskip\cmsinstskip
\textbf{Institut f\"{u}r Hochenergiephysik der OeAW,  Wien,  Austria}\\*[0pt]
W.~Adam, T.~Bergauer, M.~Dragicevic, J.~Er\"{o}, M.~Friedl, R.~Fr\"{u}hwirth\cmsAuthorMark{1}, V.M.~Ghete, C.~Hartl, N.~H\"{o}rmann, J.~Hrubec, M.~Jeitler\cmsAuthorMark{1}, W.~Kiesenhofer, V.~Kn\"{u}nz, M.~Krammer\cmsAuthorMark{1}, I.~Kr\"{a}tschmer, D.~Liko, I.~Mikulec, D.~Rabady\cmsAuthorMark{2}, B.~Rahbaran, H.~Rohringer, R.~Sch\"{o}fbeck, J.~Strauss, W.~Treberer-Treberspurg, W.~Waltenberger, C.-E.~Wulz\cmsAuthorMark{1}
% ...
\vskip\cmsinstskip
% ...
1:~~Also at Vienna University of Technology, Vienna, Austria\\
2:~~Also at CERN, European Organization for Nuclear Research, Geneva, Switzerland\\
% ...

\end{sloppypar}
\end{document}